\documentclass[12pt]{article}
\usepackage{psfrag,epsfig}
\usepackage{a4,isolatin1,amsmath}
\begin{document}
\begin{center}
{\bf \Large Few-body approach to diffraction of small helium clusters by
  nanostructures\footnote{Invited talk delivered at the
16TH EUROPEAN CONFERENCE ON FEW-BODY PROBLEMS IN PHYSICS,  Autrans,
June 1-6, 1998} 
\normalsize}

\vspace*{0.5cm}

\noindent {\bf Gerhard C. Hegerfeldt\footnote{ e-mail:
    hegerf@theorie.physik.uni-goettingen.de}}\\
Institut f\"ur Theoretische Physik, Universit\"at G\"ottingen\\
 Bunsenstr. 9, 37073 G\"ottingen, Germany\\[.2cm]
and\\[.2cm]
{\bf Thorsten K\"ohler\footnote{e-mail:
  koehler@theorie.physik.uni-goettingen.de}}\\
Max-Planck-Institut f\"ur Str\"omungsforschung\\
  Bunsenstr. 10, D-37073 G\"ottingen, Germany
\end{center}

\begin{abstract} We use few-body methods to investigate the diffraction of
weakly bound systems by a transmission grating. For helium
dimers, He$_2$, we  obtain explicit expressions for the transition
amplitude in the elastic channel.
\end{abstract}
\section{1. Introduction}
In recent years typical optics experiments have been carried over to
atoms, among them diffraction by  double slits 
\cite{firstExpMlynekPrichard} and by  transmission gratings which were
produced by nanostructure techniques
\cite{SchoellToennScience,PrichardMolecule}.
 The usual wave-theoretical
methods of classical optics (Huygens, Kirchhoff) seem to give excellent
descriptions of these experiments \cite{AtOptReviews} so that
apparently not much
more than a good textbook on classical optics seems to be needed.

This simple picture, however, has changed. Recently  Sch\"ollkopf and Toennies
 \cite{SchoellToennScience,SchoellToennJCP,tbp} have performed
 impressive diffraction  experiments  with helium dimers and helium
clusters consisting of up to 26 atoms. A typical diffraction pattern
is shown in Fig. 1. The helium dimer He$_2$, discovered a few 
years ago \cite{Luofirstdimer,SchoellToennScience}, has an extremely
low binding energy of $-0.11~ \mu {\rm eV}$ \cite{TangToenniesYiu},
with no excited states. The binding energy is much smaller than the
incident kinetic energy of typical experiments. Its diameter is
estimated to be about 6 nm \cite{LuoGieseGentry}. Present day slit
widths are as low as 50 nm and a further reduction is expected.

When a system is observed at a nonzero diffraction angle it clearly has
received a lateral moment transfer from the grating. For weakly bound
systems like He$_2$ or higher clusters this might induce breakup
processes which in turn might change the diffraction
pattern. Similarly, the size of the bound system may have an
influence. If so, what will be the effect? Will the diffraction
pattern change drastically? Will the diffraction peaks
still be at the same locations as for point particles? These
questions will be investigated and answered in the following, with
emphasis on He$_2$. More technical details can be found in Ref. 
\cite{HeKoe1}. 

Diffraction by a transmission grating really means scattering by the
bars of the grating, and to include breakup processes and finite-size
effects we use multichannel scattering theory. We describe the effect
of the grating bars by means of a short-range repulsive (`reflective')
potential. The possible role of attractive parts to the potential will
be discussed at the end. As usual we neglect the electronic degrees of
freedom.

We thus consider a weakly bound system of {\em two} particles and its
scattering by an {\em external potential}. Due to the presence of 
the external potential this problem cannot be reduced to a one-particle
problem but has more resemblance, at least mathematically,
to a three-body problem. In fact, we will show that the Faddeev
approach to the three-body problem in its formulation by Alt, Grassberger, and
Sandhas \cite{AGS} can be carried over and adapted to the two-body
problem with external potential.
\begin{figure}[hbt]
\epsfig{file=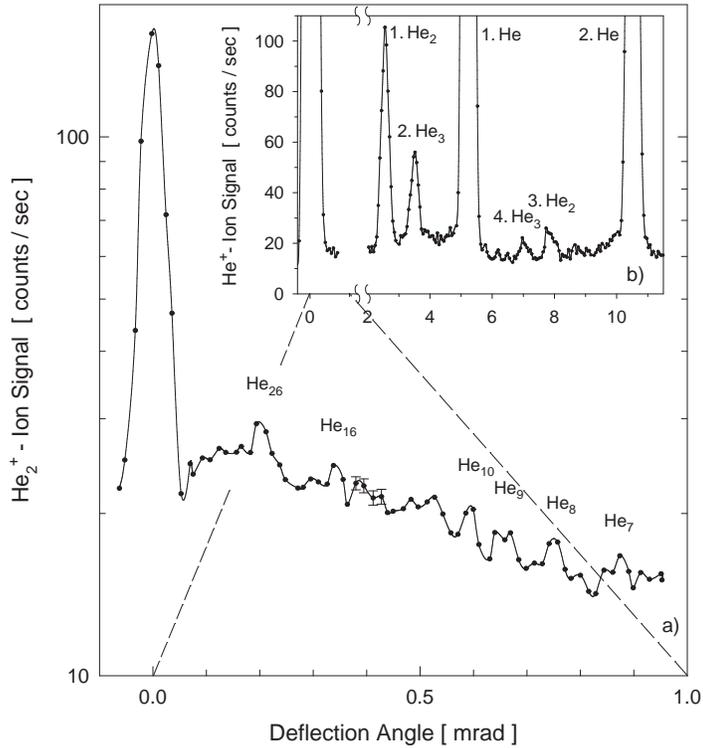,width=300pt}
\caption[]{Diffraction of helium clusters for a source
  temperature of $6\,\mbox{K}$ by a 100-nm-period transmission grating
 \cite{tbp}.}
\end{figure}
\section{ AGS Equations for  Two Particles with External Potential}
In this section we consider a bound two particle system (`molecule')
scattered by a quite general obstacle. The coordinates are ${\bf x}_1$
and ${\bf x}_2$, the binding potential is $V({\bf x}_1 -{\bf x}_2 )$,
with negative binding energies $E_\gamma$ and bound states
$\phi_\gamma({\bf x})$. The external (obstacle) potential is
\begin{equation}
  \label{a}
  W({\bf x}_1,{\bf x}_2)=W_1({\bf x}_1)+W_2({\bf x}_2)
\end{equation}
where $W_1$ and $W_2$ are short-range and strongly repulsive,
representing Dirichlet boundary conditions in the limit. For dimers
one has $W_1 = W_2$.

We first consider the process $|{\bf P}^\prime,\phi_{\gamma^\prime}
  \rangle~~\rightarrow |{\bf P},\phi_{\gamma}\rangle $
 from an incoming molecule of
  momentum ${\bf P}^\prime$ and internal state $\phi_{\gamma^\prime}$
    to an outgoing molecule of momentum ${\bf P}$ and internal state 
$\phi_{\gamma}$, i.e. no breakup. We introduce the six-dimensional
Green's operators $G_0(z) = (z-H_0)^{-1}$,
$G(z) = (z-H_0-V-W)^{-1}$, $G_V(z) = (z-H_0-V)^{-1}$,
and similarly $G_W(z)$, where $H_0$ is the free Hamiltonian (kinetic
energy). One has the usual resolvent equations
\begin{eqnarray}
  \label{c}
  &G=G_V+G_V WG\\
  \label{d}
  &G=G_W+G_W VG.
\end{eqnarray}
The $T$ matrices for $V$ and $W$ are defined as 
\begin{eqnarray}
  \label{5}
  T_V(z)&\equiv&V+VG_V(z)V\\
  \nonumber
  T_W(z)&\equiv&W+WG_W(z)W~.
\end{eqnarray}
Writing 
\begin{equation}
  T_V=VG_V(G_V^{-1}+V)=VG_VG_0^{-1}
\end{equation}
and similarly for $W$ gives
\begin{eqnarray}
  \label{17}
  &T_VG_0=VG_V\\
  \nonumber
  &T_WG_0=WG_W~.
\end{eqnarray}
We now introduce operators $U_{VV}(z)$ and $U_{WV}(z)$ through
\begin{eqnarray}
  \label{e}
  G(z)&=&G_V(z)+G_V(z)U_{VV}(z)G_V(z)\\
  \label{f}
  G(z)&=&\ \ \ \ \ \ \ \ \ \ \ \ G_W(z)U_{WV}(z)G_V(z).
\end{eqnarray}
As in Ref. \cite{AGS} one easily shows that the transition amplitude
for the above process is
\begin{equation}
  \label{12}
  t({\bf P},\phi_\gamma;{\bf P}^\prime,\phi_{\gamma^\prime})=
  \langle{\bf P},\phi_\gamma|U_{VV}(E+{\rm i}0)
  |{\bf P}^\prime,\phi_{\gamma^\prime}\rangle~.
\end{equation}
Equating Eqs. (\ref{c}) and (\ref{e}), 
\begin{equation}
  G_V+G_VWG=G_V+G_VU_{VV}G_V,
\end{equation}
and inserting Eq. (\ref{f}) gives 
\begin{equation}
  \label{h}
  U_{VV}=WG_WU_{WV}=T_WG_0U_{WV}
\end{equation}
Similarly, equating Eqs. (\ref{d}) and (\ref{f}) and
inserting Eq. (\ref{e}) gives
\begin{equation}
  \label{i}
  U_{WV}=G_0^{-1}+T_VG_0U_{VV}.
\end{equation}

The last two equations are the AGS equations for two particles
interacting through a potential $V$ and with an additional external
potential $W$. These AGS equations decouple here, as seen by insertion of
Eq. (\ref{i}) into Eq. (\ref{h}), which gives 
\begin{equation}
  \label{18}
  U_{VV}=T_W+T_WG_0T_VG_0U_{VV}.
\end{equation}

This equation is exact and contains both breakup and finite-size
effects. It lends itself to iteration, and to lowest order in $T_V$ one has 
\begin{equation}
  U_{VV}\cong T_W~.
\end{equation}
By Eq. (\ref{12}) this gives
\begin{equation}
  \label{21}
  t({\bf P},\phi_\gamma ;{\bf P}^\prime,\phi_{\gamma^\prime})
  \cong \langle {\bf P},\phi_\gamma |T_W(E+{\rm i}0)
  |{\bf P}^\prime,\phi_{\gamma^\prime}\rangle.
\end{equation}
Although $T_W$ is the transition operator for scattering of two 
asymptotically free
particles by the external potential $W$ it is not on the energy shell
in Eq. (\ref{21}). Here the interaction $V$ is taken into account through
the wave-functions $\phi_\gamma$ and $\phi_{\gamma'}$.

Similarly to Eq. (\ref{12}) one can show that 
\begin{equation}
  \label{19}
  U_{0V}\equiv G_0^{-1}+T_VG_0U_{VV}+T_WG_0U_{WV}
\end{equation}
is the transition operator for breakup processes,
and for this one only has to know $U_{VV}$ and $U_{WV}$.

\section{Application to Diffraction of Helium Dimers}
The scattering amplitude in Eq. (\ref{21}) can be evaluated in case of
small diffraction angles for a
transmission grating of $N$ bars of period $d$ and slit width $s$
\cite{HeKoe1}. We consider  a bound system of two particles, each of
mass $m$, and normal incidence. With $P$ the total momentum
of the system and $P_2$ the
lateral momentum transfer, the elastic scattering amplitude becomes of
the form
\begin{equation}
  \label{38}
  t({\bf P},\phi_\gamma;{\bf P}^\prime,\phi_\gamma)
  =t_{\rm coh}(P_2)+t_{\rm incoh}(P_2).
\end{equation}
The first term is a superposition of amplitudes from individual bars,
while the second results from interactions with separate bars
and it can be shown to be {\em negligible} if
$d-s$ and $s$ are larger than the molecule. The coherent part
is of the form
\begin{equation}
  \label{39}
  t_{\rm coh}(P_2)=t^{\rm mol}_{\rm bar}(\gamma,P/M;P_2)
    \frac{\sin(P_2 N d/2\hbar)}{\sin(P_2 d/2\hbar)}  \delta(P_3)
\end{equation}
where the first factor is the molecular transition amplitude for a
single bar of width $d - s$, for total momentum $P$ and total mass
$M$.  The second factor is the usual sharply
peaked grating function known from optics, and it gives the
diffraction peaks at the same locations as for
point-particles with corresponding lateral momentum transfer. The
$\delta$ function is due to the assumed infinite extent of the grating
in the 3 direction. The amplitude of
the diffraction  peaks is determined by $t^{\rm mol}_{\rm bar}$. The
latter has been 
explicitly calculated by means of Eq. (\ref{21}) in Ref. \cite{HeKoe1}
and it depends only on the absolute value of the ground-state
wave-function. One obtains 
\begin{eqnarray}
  \label{40}
&\!\!\!\!\!\!\!\!\!\!\!\!\!\!\!\!\!\!\!\!\!\!\!\!\!\!\!\! t^{\rm mol}_{\rm bar}(\gamma,P/M;P_2)
 =-\frac{2{\rm i} P}{(2\pi)^2 M}
  \frac{\sin[P_2(d-s)/2\hbar]}{P_2}
  \int{\rm d}^3 x
      \,2\,{\rm e}^{{\rm i} P_2 x_2/2 \hbar}
  |\phi_\gamma({\bf x})|^2\\
  \nonumber
  &~~~~~~~~~~~~~~~+\frac{4 {\rm i} P}{(2\pi)^2 M}
  \int{\rm d} x_1 {\rm d} x_3 
  \int_0^{d-s} {\rm d} x_2
  |\phi_\gamma({\bf x})|^2
    \sin
    \left[
      \frac{P_2}{2\hbar}(d-s-x_2)
    \right]
      /P_2~.
\end{eqnarray}
The expression preceding the first integral is the single-bar amplitude for a
point particle of mass M and total momentum $P$. The complete expression on the
r.h.s. reduces to this if $|\phi_\gamma({\bf x})|^2$ contracts to a point.

In Fig. 2 we have plotted $|t^{\rm mol}_{\rm bar}|^2$ for He$_2$ and the
corresponding quantity for a point particle. The bar width is 25 nm.
For a symmetric grating of period 50 nm the vertical lines in the inset 
indicate where the peaks of the grating functions cut out the diffraction
peaks of first order, second order and so on. For a symmetric grating
and point particles there are no 
even-order diffraction peaks since the single-bar amplitude vanishes
there, as  seen in  the inset. For the dimer this is not true, and it 
 therefore has small even-order peaks which become more pronounced for
 smaller bar and slit widths.

\begin{figure}[hbt]
\epsfig{file=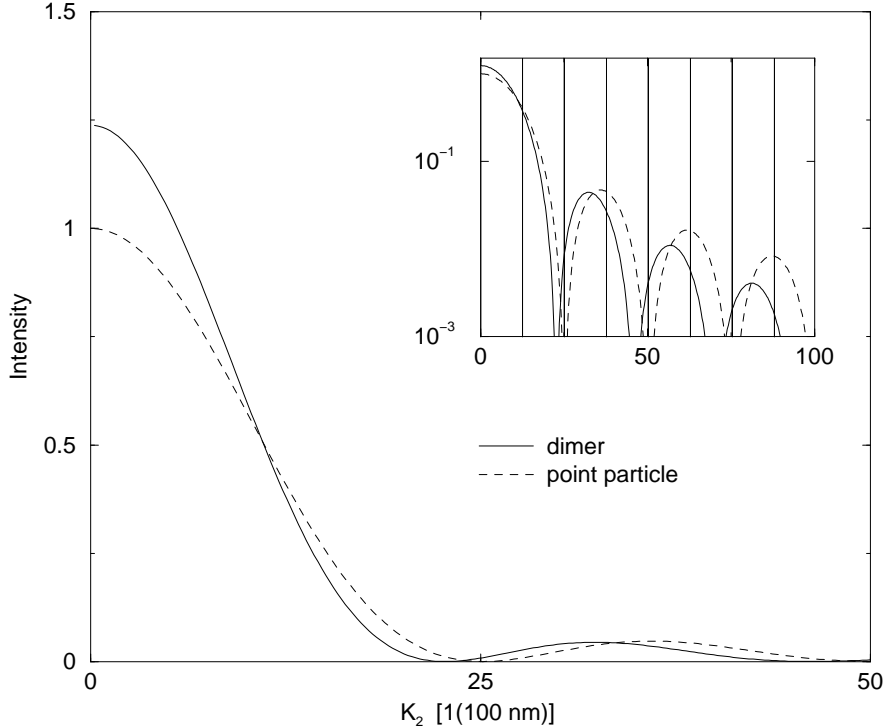,width=280pt,angle=270}
\caption[]{Diffraction by a single bar of width 25 nm for helium dimers and 
   point particles. For a symmetric grating with this bar and slit width
   the vertical lines in the inset pick out the corresponding
   diffraction peaks. For dimers there are even order peaks which
   are absent for point particles. The lateral momentum transfer is
   $P_2 =\hbar K_2$.} 
\end{figure}

Intuitively we expect that finite-size effects of He$_2$ might be
partially taken into account by a larger bar and smaller slit
width. Fig. 3 shows that the single-bar amplitude of He$_2$, for a bar
width of 25~nm, can be approximated for small lateral momentum transfer by
that of a point particle for bar width of (25 + 2.8)~nm. For a grating
this would mean a nonsymmetric grating of correspondingly smaller slit
width. The expectation value of $|x_2|$ for the ground-state
wave-function of He$_2$ we are using is just 2.8~nm. A cumulant
expansion of $t^{\rm mol}_{\rm bar}$ shows that the equality of these
two numbers  is not a coincidence \cite{HeKoe2}.
\begin{figure}[hbt]
\epsfig{file=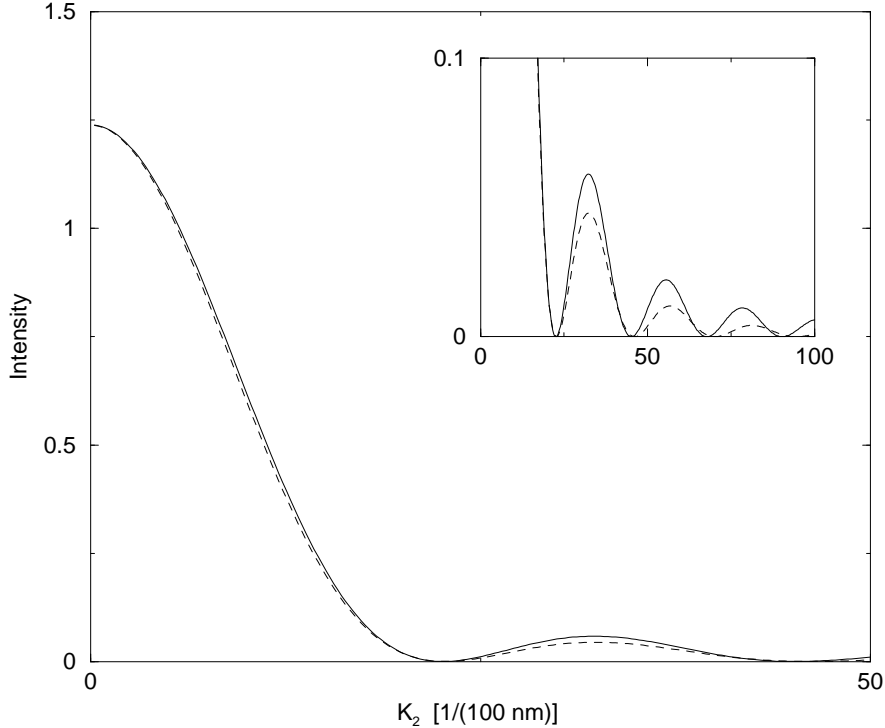,width=280pt,angle=270}
\caption[]{Diffraction of helium dimers (solid line) by a single bar
  of width 25 nm and diffraction of point particles (dashed line) by a
  single bar of width (25 + 2.8) nm. There is agreement for small
  lateral momentum   transfer $P_2 =\hbar K_2$.} 
\end{figure}

\section{Location of the Diffraction Peaks}
Can one understand more directly and intuitively why the diffraction
peaks occur at the same locations as for point particles of the same
lateral momentum transfer? This is easy to see if one accepts that
$t_{\rm incoh}$ in Eq. (\ref{38}) is negligible and that one just has
to add the amplitudes from the individual bars. Indeed, the potential
of the $n$th bar is the translate
\begin{equation}
\label{translate}
\exp\{-{\rm i}n\hat{P}_2d/\hbar\}W_{\rm bar}
  \exp\{{\rm i}n\hat{P}_2d/\hbar\}
\end{equation}
of the first bar potential, and similarly for the $n$th bar
transition amplitude $U_{VV}^{(n)}$. Since $|{\bf
  P},\phi_{\gamma}\rangle$ is an eigenvector of the momentum operator
$\hat P_2$, the associated amplitude differs just by the phase factor 
 $\exp\{-{\rm i}n{P}_2d/\hbar\}$ from that of the first bar. Summation
 over $n$ leads to a sum over the phase factors, resulting in the
 grating function and Eq. (\ref{39}).

There is also a symmetry argument for an infinite grating, exact to
all orders and holding for larger clusters, too. For an infinite
grating the complete Hamiltonian has period $d$ under translations in
the 2 direction. Hence the Hamiltonian as well as the associated
time-development operator $U(t)$ commute with $\exp\{-{\rm
  i} \hat{P}_2d/\hbar\}$. As a consequence 
$$
\langle {\bf P},\phi_\gamma |U(t)
  |{\bf P}^\prime,\phi_{\gamma^\prime}\rangle
$$
must vanish unless 
\begin{equation}
\label{dir}
P_2/\hbar = \frac{2\pi}{d} n~, ~~~~~~~n=0,\pm 1, \cdots
\end{equation}
and thus there are only transitions in these directions.

We recall that
experimentally one always has a certain momentum spread and that only
a part of the grating is illuminated by the incoming beam. To take
this into account one can modify the above plane-wave formulation
 and it allows one to let $N$ become arbitrarily large \cite{HeKoe2}. The
momentum spread then just leads to a broadening of the diffraction
peaks. Incidentally,  this procedure yields  the
correct height for the zeroth-order peak
which otherwise would be misrepresented by a finite grating illuminated by an
infinite plane wave. Figs. 4 and 5 show some results.
\begin{figure}[hbt]
\epsfig{file=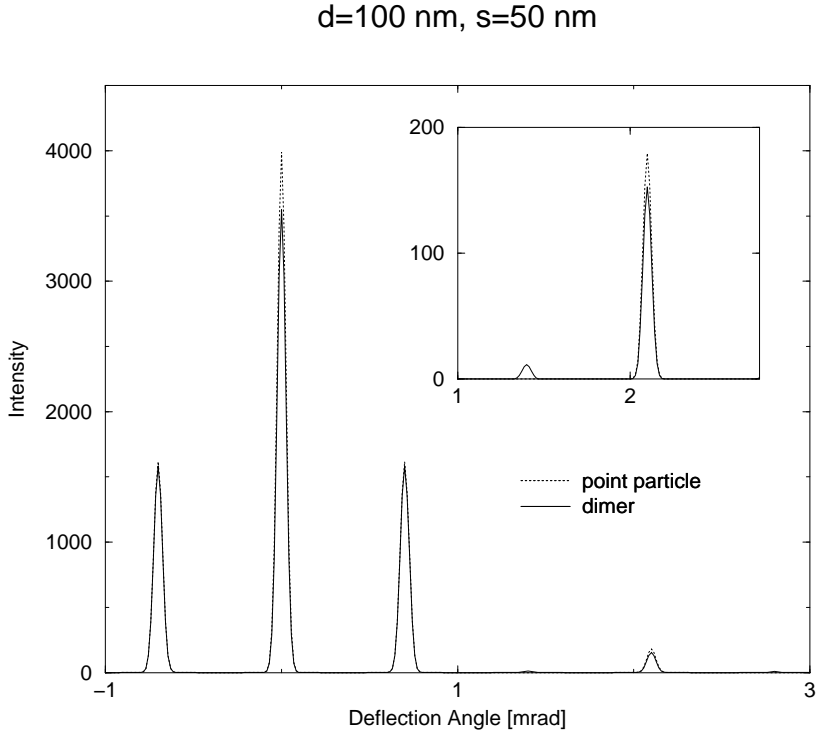,width=280pt,angle=270}
\caption[]{Diffraction  by a symmetric grating
  with slit width 50 nm for helium dimers (solid line) and for
  corresponding point  particles (dotted line). The differences are
  small, the second order peak barely visible.}
\end{figure}
\begin{figure}[hbt]
\epsfig{file=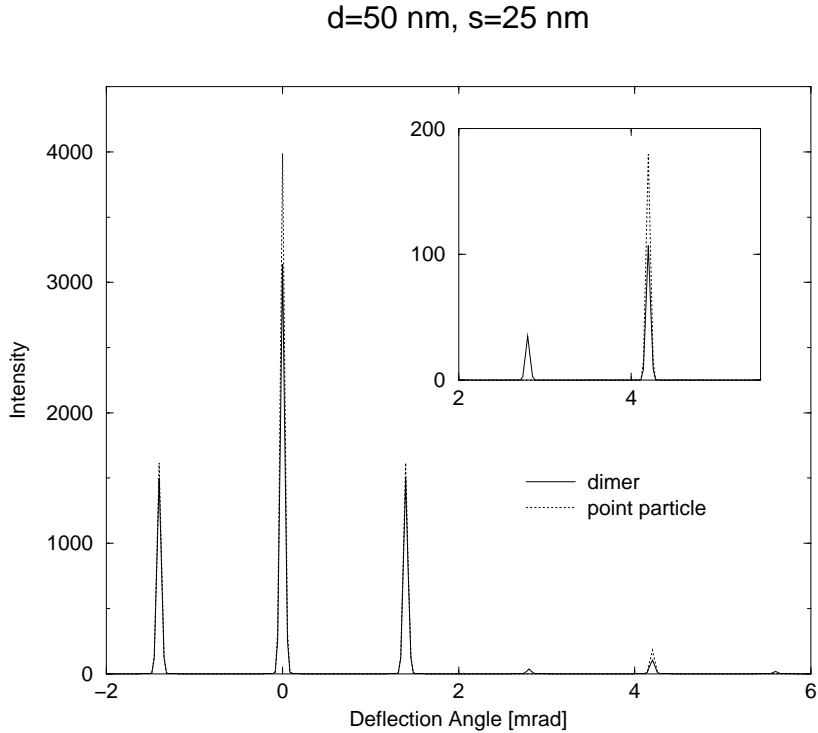,width=280pt,angle=270}
\caption[]{Diffraction  by a symmetric grating
  with slit width 25 nm for helium dimers (solid line) and for point
  particles (dotted line). The differences are more pronounced, the
  second order peak is larger.}
\end{figure}

\section{Further Developments}
Recent experiments by Sch\"ollkopf and Toennies \cite{priv1} have shown
that an additional attractive potential may have an unexpectedly large
influence on the diffraction patterns for higher noble atoms,
increasing from neon to argon to krypton, in line with the
increasing polarizability of these atoms. This attractive potential will
also influence the diffraction pattern of bound systems to some extent.

In a first step we have taken this additional surface potential into account
for diffraction of atoms. This can be done by using similar resolvent 
methods as
in Section 2, with the surface potential treated as a perturbation. In the
course of this something interesting happened. At first our results
did not even qualitatively agree with the experiments for higher noble
atoms -- in particular the hierarchy of the peaks did not come out
correctly -- until we learned that due to the etching process the
cross section of the grating bars was not rectangular as assumed by
us but rather trapeze-like, with the angle off the perpendicular by
$8^0$. Using this geometry agreement with the experiments was then
obtained \cite{HeKoe2}. This, incidentally, shows the sensitivity of
the theoretical methods.

The next step is to take the attractive surface potential into account
for helium dimers. This work is presently in progress.

Another interesting topic is to use the cumulant expansion mentioned
in the last section to obtain some measure of the `size' of He$_2$ from
diffraction data and to obtain consequences for the binding potential.

In principle it should be possible to carry the approach of
Section 2 over to three particles with external potential and apply it
to the helium trimer, He$_3$. In this connection a very interesting
question is whether one can draw any conclusions from He$_3$
diffraction patterns for the theoretically predicted Efimov state of
He$_3$ \cite{Ef1,Ef2}.
\section{Conclusions}
 We have used few-body methods to investigate the diffraction of
 weakly bound two-particle 
systems by a transmission grating and have obtained explicit
expressions for the transition amplitude in the elastic
channel.
The diffraction pattern is a product of
the single-bar amplitude $t^{\rm mol}_{\rm bar}$  and the usual grating
function. As for point particles our result depends only on $P_2/\hbar =
(2 \pi/\lambda)\sin\varphi$ and on the bar width. The
zeros of $t^{\rm mol}_{\rm bar}$ may differ from those for point particles so
that even-order diffraction peaks may appear for symmetric gratings.

For large bar and slit widths ($\sim10~\times$ bound-state diameter)
the diffraction 
pattern is still very close to that for point particles, although minor
deviations exist (second order peak). 
For smaller bar and slit widths ($\sim5~\times$ diameter) significant
deviations occur. The third order peak can be up to 40\% lower. This
can be attributed to breakups which should increase for larger
momentum transfer. There is also an
appreciable second order peak which is absent for point particles. This is
more difficult to understand by breakups and is rather a finite-size
effect since the finite size can be partially taken into account by
considering point particles and a
nonsymmetric grating with smaller slit widths.

Additional attractive surface potentials which can, and do, modify the
diffraction patterns can be taken into account.

Extension to helium trimers seems feasible, with a possible connection
to Efimov states.\\

{\em Acknowledgments:} We would like to thank W. Sch\"ollkopf and
  J. P. Toennies for letting us  use some of their as yet
  unpublished results.

 \end{document}